\documentclass[superscriptaddress,twocolumn,floatfix]{revtex4}
\usepackage{graphicx}
\usepackage{amsmath}
\usepackage{color}
\usepackage{ulem}

\begin{document}

\title{Modulating light with light via giant nano-opto-mechanical nonlinearity of plasmonic metamaterial}

\author{Jun-Yu Ou}
\affiliation{Optoelectronics Research Centre and Centre for Photonic Metamaterials, University of Southampton, SO17 1BJ, UK}

\author{Eric Plum}
\email{erp@orc.soton.ac.uk}
\affiliation{Optoelectronics Research Centre and Centre for Photonic Metamaterials, University of Southampton, SO17 1BJ, UK}

\author{Jianfa Zhang}
\affiliation{Optoelectronics Research Centre and Centre for Photonic Metamaterials, University of Southampton, SO17 1BJ, UK}
\affiliation{College of Optoelectronic Science and Engineering, National University of Defense Technology, Changsha, 410073, China}

\author{Nikolay I. Zheludev}
\email{niz@orc.soton.ac.uk}
\homepage{www.nanophotonics.org.uk}
\affiliation{Optoelectronics Research Centre and Centre for Photonic Metamaterials, University of Southampton, SO17 1BJ, UK}
\affiliation{The Photonics Institute and Centre for Disruptive Photonic Technologies, Nanyang Technological University, Singapore 637378, Singapore}

\date{\today}

\begin{abstract}
From the demonstration of saturable absorption by Vavilow and Levshin in 1926 \cite{Wawilow_1926}, and with invention of the laser, unavailability of strongly nonlinear materials was a key obstacle for developing optical signal processing, in particular in transparent telecommunication networks. Today, most advanced photonic switching materials exploit gain dynamics \cite{Science_NonlinearGain_1999} and near-band and excitonic effects in semiconductors \cite{Nature_UltrafastLasers_SESAMs_2003}, nonlinearities in organic media with weakly-localized electrons \cite{Book_Nonlinear_Organics_2001} and nonlinearities enhanced by hybridization with metamaterials \cite{NonlinearMetamaterialsReview_2014}.
Here we report on a new type of artificial nonlinearity that is nano-opto-mechanical in nature. It was observed in an artificial metamaterial array of plasmonic meta-molecules supported by a flexible nano-membrane. Here nonlinearity is underpinned by the reversible reconfiguration of its structure induced by light. In a film of only 100 nanometres thickness we demonstrated modulation of light with light using milliwatt power telecom diode lasers.
\end{abstract}

\maketitle

\begin{figure} [tbh]
\includegraphics[width=80mm]{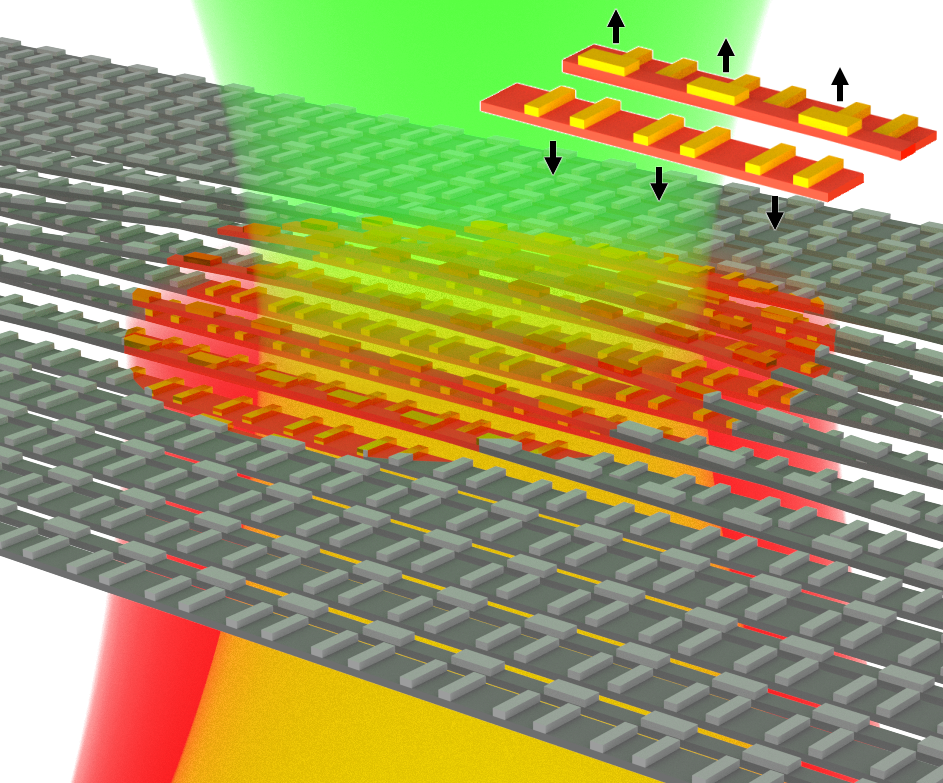}
\caption{\label{fig-concept}
\textbf{Nonlinearity in nano-opto-mechanical metamaterial.}
Light-induced (red) forces in the plasmonic metamaterial array cause nanoscale reversible displacements of its small and light building blocks supported by an elastic nano-membrane. They move fast, with a response time reaching microseconds. These displacements change the plasmonic spectra of the metamaterial array and its transmission. This is used to modulate another, weaker beam of light (green) at a different wavelength.
}
\end{figure}

Some exceptional opportunities for developing engineered nonlinear media are provided by nano-optomechanics that take advantage of the changing balance of forces at the nanoscale \cite{QuantumOptomechanicsReview_2012, Nature_NanomechanicalGroundState_2011, Nature_QuantumOptomech_2010, Phystoday_2007_lamoreaux_casimir, Phystoday_2004_Fitzgerald_spinstate, NL_Polman_Optomech_2013, AdvMater_2013_NanoPillarSensor_Zayats, ACSphoton_2014_ParallelOptoMech_Ewold}. With the decrease in the physical dimensions of a system the electromagnetic forces between constituent elements grow, as may be illustrated by the repulsion of electrons as their separation diminishes. In contrast, elastic forces, such as the force restoring a compressed spring, decrease with size. Moreover, the nanoscale metamaterial building blocks have high natural frequencies and thus can be moved very fast, potentially offering GHz switching bandwidth for elements of sub-micron size.

Recently this was exploited to develop reconfigurable metamaterials fabricated on nanoscale elastic membranes. These photonic metamaterials can be driven thermo-elastically \cite{ThermalRPM} and with electromagnetic forces, such as the Coulomb force between charged elements of the nanostructure \cite{ElectroRPM}, or the Lorentz force acting on currents running through conductive elements in magnetic field \cite{NatComms_2015_MEO_RPM}. Their electro-optical and magneto-optical switching characteristics surpass those of natural media by orders of magnitude.

Metamaterials are in essence arrays of optical resonators where light-induced nano-opto-mechanical phenomena can play crucial roles. For instance it has been suggested that nearfield \cite{PRB_2012_GeckoToeMM_Numerical} and Casimir forces \cite{PRL_2008_Rosa_casimirMM, PRL_2009_Zhao_CasimirMM} in metamaterials are modified in the presence of light. Forces between oscillating plasmonic or displacement currents induced by light in metamolecules were also theoretically shown to be sufficient to drive reconfiguration of a metamaterial structure in the optical part of the spectrum \cite{LightSciAppl_2013_DielectricOptomechMM_Numerical, OE_2010_OptForce_NanowirePair_Numerical, OL_DielectricMMOptForces_2014}, and proof-of-principle experiments on such interactions between individual metamolecules have been reported at microwave frequencies \cite{NatMater_2012_MagnetoelasticMM}.

Here we show that reversible reconfiguration of a plasmonic metamaterial nanostructure driven by optical forces between its illuminated elements can be the source of a strong optical nonlinearity. We show that this cubic nonlinearity may be used to modulate light with light at milliwatt power levels. Moreover, although for the majority of media the magnitude of nonlinearity tends to be proportional to its response time, the nano-opto-mechnical nonlinearity is three orders of magnitude faster than could be expected from this otherwise universal trend.

The plasmonic nanomechanical metamaterial was fabricated on strips of suspended dielectric membrane of nanoscale thickness in such a way that plasmonic elements of the metamolecules were located on different strips. Here light illumination leads to electromagnetic and thermal forces that induce nanoscale strip displacements and thus reconfiguration of individual metamolecules in a way that affects their plasmonic resonances, see Fig.~\ref{fig-concept}. This leads to modulation of the metamaterial's transmissivity at different wavelengths. The nonlinear response of metamaterial, that is only 100 nanometres thick, can be observed at only a few milliwatts of continuous laser power, using lasers operating at telecommunication wavelengths.

\begin{figure} [tbh]
\includegraphics[width=80mm]{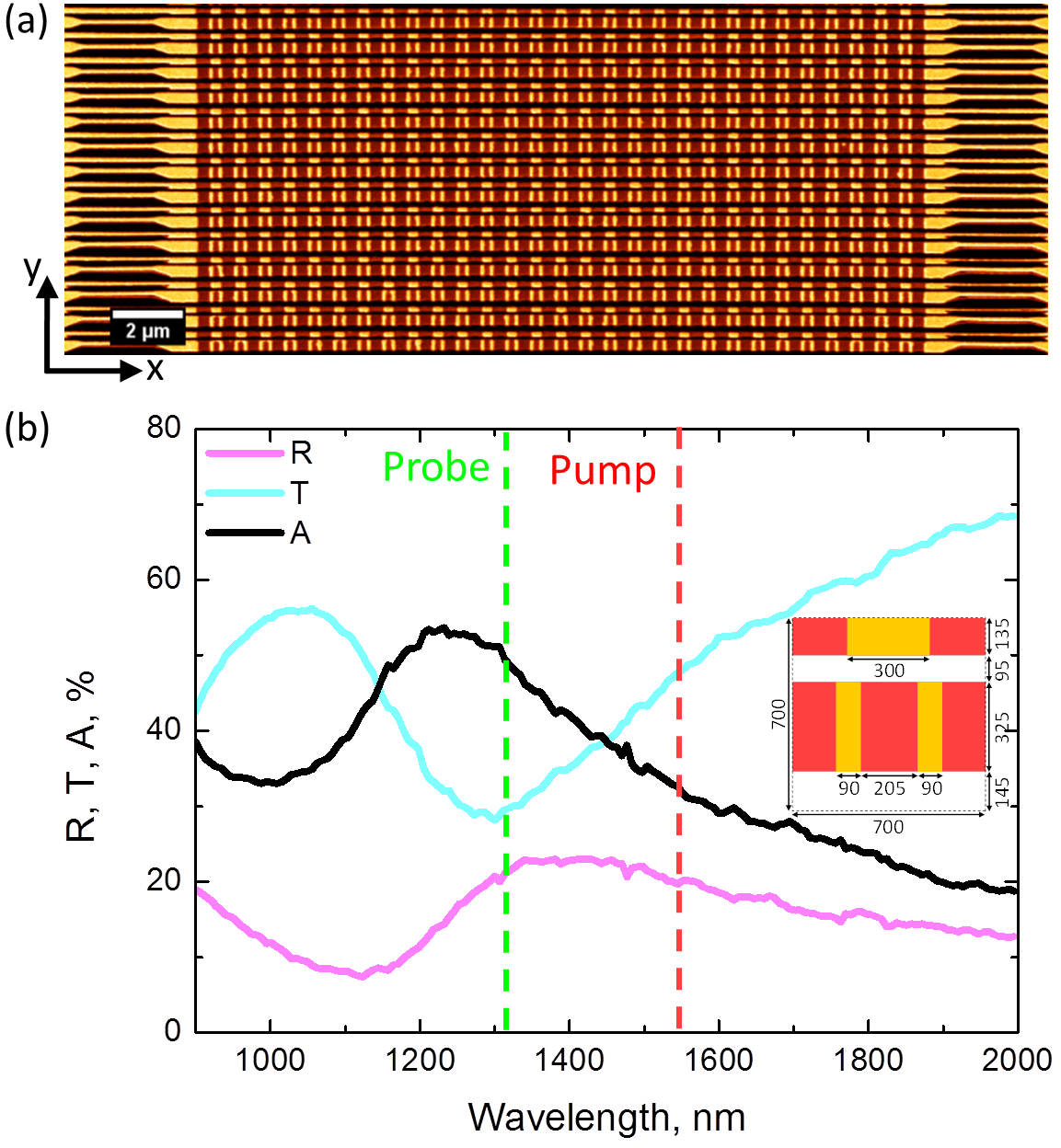}
\caption{\label{fig-sample}
\textbf{Optically reconfigurable plasmonic metamaterial.} (a) Scanning electron microscope image of the metamaterial nanostructure consisting of asymmetrically spaced gold (yellow) resonators on silicon nitride strips (red).
(b)~Transmission $T$, reflection $R$ and absorption $A$ spectra of the metamaterial for $x$-polarized light incident on the silicon nitride side. Dashed lines indicate the optical 1310~nm probe and 1550~nm pump wavelengths used in the modulation measurements of the main manuscript. The inset shows detailed unit cell dimensions in nanometers. The scanning electron micrograph was recorded by FEI Helios 600 NanoLab and the optical spectra were measured using a microspectrophotometer (CRAIC Technologies).
}
\end{figure}

We developed nano-opto-mechanical metamaterial on the basis of a $\Pi$-shaped resonator design known for exhibiting plasmon-induced transparency \cite{PRL_2008_PlasmonEIT,OPN_2009_MM_EIT}, see Fig.~\ref{fig-concept} and Fig.~\ref{fig-sample}a. To allow mechanical deformation of the 700~nm $\times$ 700~nm plasmonic $\Pi$ metamolecules, the horizontal and vertical gold bars were supported by different flexible silicon nitride strips of 28~$\mu$m length spaced by alternating gaps of 95~nm and 145~nm. The nanostructure was fabricated by focused ion beam milling from a 50~nm thick silicon nitride membrane covered by 50~nm of gold. Such structures can be reconfigured by (i) differential thermal expansion through light-induce heating of the bimorph gold and silicon nitride layers and (ii) by near-field optical forces acting between elements of illuminated plasmonic resonators.

\begin{figure}
\includegraphics[width=80mm]{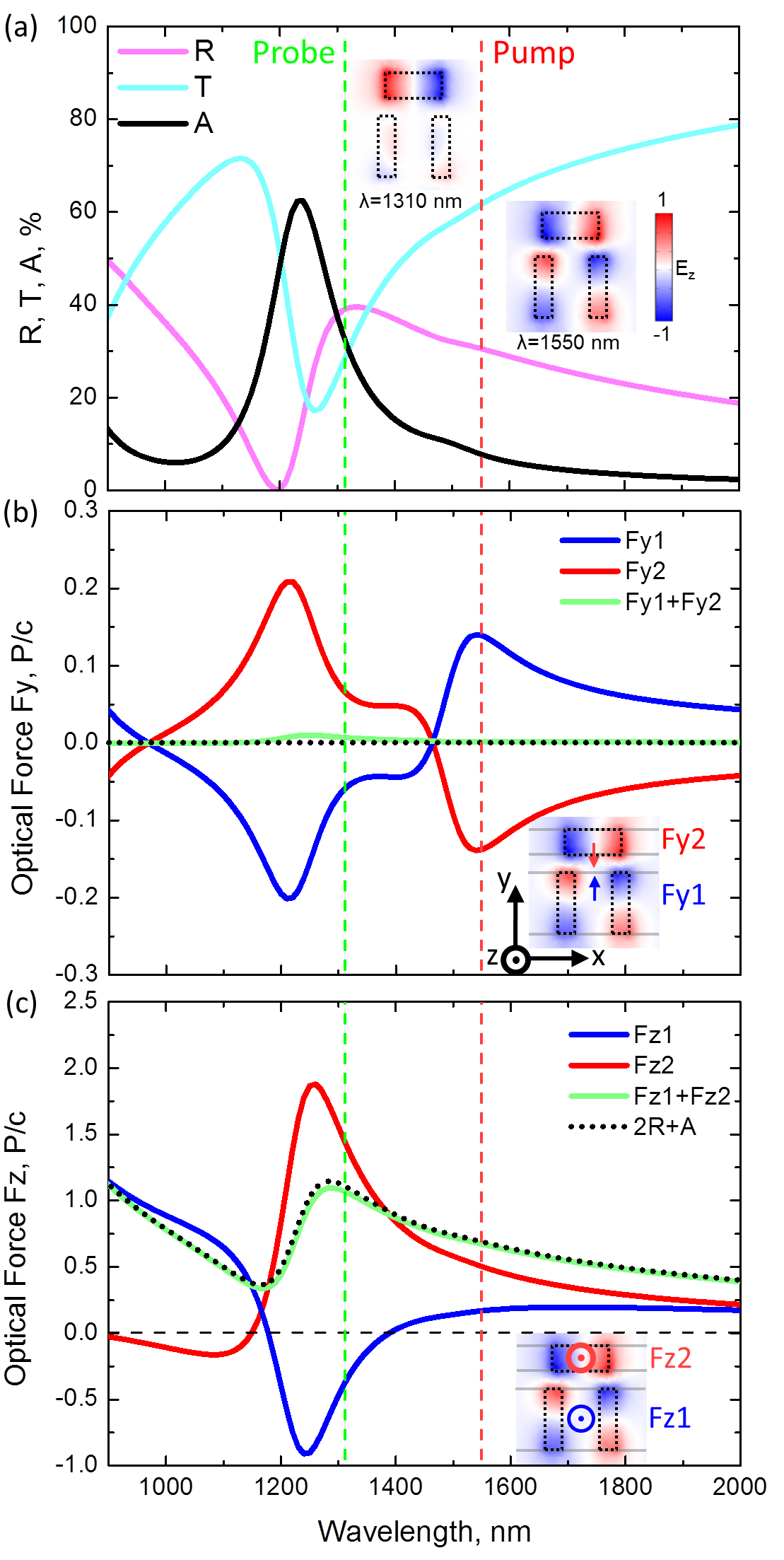}
\caption{\label{fig-simulations}
\textbf{Optical spectra and optical forces.} (a) Simulated metamaterial transmission $T$, reflection $R$ and absorption $A$ spectra. Insets show maps of the instantaneous values of the optically induced charges at the probe and pump wavelengths.
(b)~In-plane-of-metamaterial component of optical forces $F_{y1,2}$ acting between the strip segments of an individual unit cell according to Maxwell stress tensor calculations.
(c)~Normal to the metamaterial component of optical forces $F_{z1,2}$ acting on the strip segments along the light propagation direction.
The total optical force on the unit cell (green line) is presented alongside the value expected from reflection and absorption (dotted black line). Forces are shown per unit cell in units of $P/c$, where $P$ is the incident power per unit cell and $c$ is the speed of light in vacuum. All quantities are shown for $x$-polarized illumination of the silicon nitride side.
}
\end{figure}

\begin{figure*} [t]
\includegraphics[width=159mm]{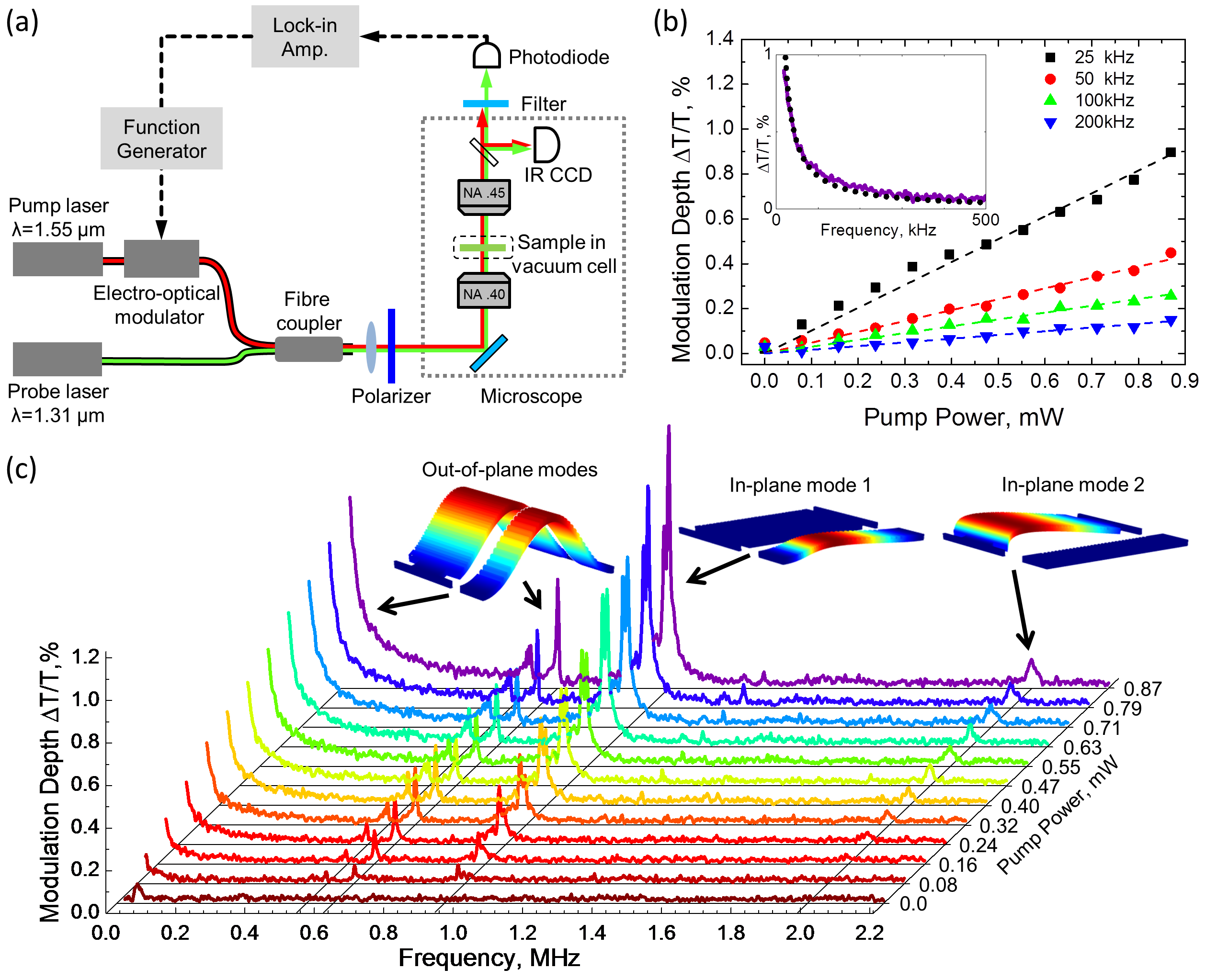}
\caption{\label{fig-modulation}
\textbf{Modulating light with light using nano-opto-mechanical metamaterial.} (a) Schematic of the pump-probe experimental setup. (b)~Transmission modulation depth at $\lambda=1310$~nm as a function of pump power for several pump modulation frequencies and pump modulation depth close to 100\%. The inset shows a hyperbolic fit $\sim 1/f$ (black dashed line) of the thermal modulation tail for a pump power of 0.9~mW. (c)~Modulation depth as a function of pump laser power and modulation frequency. Corresponding mechanical modes of the metamaterial nanostructure are shown near resonant peaks.
}
\end{figure*}

Experimental measurements and full 3D Maxwell simulations show a pronounced near-infrared absorption resonance around 1240~nm, Fig.~\ref{fig-simulations}a and Fig.~\ref{fig-sample}b.
Maxwell stress tensor calculations \cite{LightSciAppl_2013_DielectricOptomechMM_Numerical} assuming normal incidence illumination of the non-diffracting, periodic metamaterial reveal optical forces acting on the $\Pi$-resonators around these resonances, see Fig.~\ref{fig-simulations}b-c.
As the normally incident photons only carry momentum along the $z$-direction
the net force along $z$ must comply with the momentum transfer associated with absorption $A$ and reflection $R$, $F_{z1}+F_{z2}=(A+2R)P/c$, where $P$ is the incident power per unit cell area and $c$ is the speed of light in vacuum. In close agreement with these relationships, our simulations show substantial differential optical forces $F_2-F_1$ between the unit cell's strip segments reaching about 0.4~$P/c$ along $y$ and 2.8~$P/c$ along $z$.

Optical forces in such metamaterial nanostructures can be understood as time-averaged Coulomb and Lorentz forces acting between the currents and oscillating dipole charges of the plasmonic resonators in the presence of the illuminating electromagnetic wave \cite{OL_DielectricMMOptForces_2014}. For example, repulsive and attractive optical forces within the metamaterial plane for illumination at wavelengths of 1310~nm and 1550~nm correspond to repulsive and attractive interaction of the dipole charges on neighboring strips, see Fig.~\ref{fig-simulations}a,b. Forces normal to the metamaterial plane result from a combination of optical pressure and the Lorentz force acting on the moving charges of the plasmonic mode in the presence of the incident wave's magnetic field.

We investigated the nonlinear optical properties of the metamaterial in a pump-probe experiment as presented in Fig.~\ref{fig-modulation}a. As the pump and probe optical sources we used fiber-coupled telecommunication laser diodes operating at the wavelengths of 1550~nm and 1310~nm, respectively. The intensity of the pump beam was modulated by a fibre-coupled electro-optical modulator. The pump and probe were then combined into a single beam using a fiber coupler, de-coupled into free space and focused on the sample placed in a microscope using focusing and collection objectives. The sample was placed in a cell evacuated to the pressure of 30~Pa to prevent damping of the nano-mechanical motion. Light transmitted through the sample was detected by an InGaAs photodetector and a lock-in amplifier locked to the pump modulation frequency. In order to optimize the light-induced nano-opto-mechanical modulation, we pumped the nanostructure at 1550~nm just above its absorption resonance, where simulations predict significant in-plane forces.

At low modulation frequencies, below 100~kHz, the optical pump leads to pronounced modulation of the metamaterial's transmission at the probe wavelength, see Fig.~\ref{fig-modulation}c. For 0.9~mW pump power (peak intensity 920~W/cm$^2$) a modulation amplitude on the order of 1$\%$ is observed at 25~kHz modulation. The low-frequency component of the nonlinear response has a thermal nature and drops rapidly with increasing modulation frequency, fading at a few 100s of kHz. In this frequency range, the modulation amplitude halves as the modulation frequency doubles which is consistent with a thermal deformation of the bimorph Au/SiNx membrane, see inset to Fig.~\ref{fig-modulation}b.
This is consistent with thermal calculations for 50~nm gold \cite{Gold_50nm_ThermalConductivity} and silicon nitride \cite{SiliconNitride_50nm_ThermalProperties_2009} layers, which predict conductive cooling timescales on the order of 20-30~$\mu$s.
The light-induced heating of the nanostructure is balanced by heat conduction along the metamaterial strips and according to our calculations can reach hundreds of degrees with pump excitation at 1~kW/cm$^2$.
As the thermal expansion coefficient of gold (14.4 $\times 10^{-6}$/K) exceeds that of silicon nitride (2.8 $\times 10^{-6}$/K) 5-fold such large temperature changes will bend the strips.
When heated by the modulated pump laser narrow strips will rise higher than wider strips due to more complete gold coverage; the metamolecule changes its shape and therefore its plamonic absorption and transmission for the probe wavelength change.

While the thermal nonlinearity easily dominates at low modulation frequencies, other modulation mechanisms are at play at higher frequencies. Indeed, as differential thermal expansion in the layered structure can only drive motion out of the metamaterial plane, it cannot drive in-plane resonant oscillations of the metamaterial strips.
Several resonant peaks of nonlinear response are seen at much higher frequencies around 600~kHz, 1~MHz and 2~MHz, see Fig.~\ref{fig-modulation}c. Mechanical eigenmode calculations identify the resonances as the fundamental modes of oscillation of the narrow and wide strips, corresponding to vibration normal to the metamaterial plane (around 600 kHz) and motion within the metamaterial plane (at MHz frequencies), see insets. Comparing the sub-$\mu$s heating and cooling cycles to the characteristic strip cooling timescale, we argue for their non-thermal origin.

To assess the origin of the high frequency modulation we estimate the optomechanical deformation that can be expected from the optical forces acting between different parts of the metamolacular resonator. Based on the results of simulations presented in Fig.~\ref{fig-simulations}c, pump-induced optical forces reach $0.5~P/c$ ($0.2~P/c$) for the narrow (wide) strips. For a strip pair with 29 unit cells and 920~W/cm$^2$ pump intensity, this corresponds to about 220~fN (90~fN) force per strip. Estimating the spring constant $k$ from strip mass $m$ and measured resonance frequency $f$ as $k=(2\pi f)^2 m =$28 fN/pm (56 fN/pm) for out-of-plane deformation, we may expect the gap between strips and thus between elements of the metamolecule to change by 6~pm. Corresponding estimates for in-plane deformation predict a gap change on the order of 1~pm. Such modulation is not sufficient to create a detectable level of non-resonant high-frequency modulation. However, much larger displacements are expected at the structure's mechanical resonances, where the displacement will be enhanced by the quality factor of the mechanical resonator. Our optical measurements show mechanical resonance quality factors on the order of 100, which may serve as a lower limit for the quality factor of mechanical resonances of individual strips as the overall resonance profile is affected by inhomogeneous broadening due to slight structural variations from strip to strip. Taking this enhancement into account, optical forces are definitely sufficient to induce resonant strip displacements on the order of 1~nm, which is consistent with the observed transmission modulation on the order of 1\%. Therefore we argue that the resonant response at 600 kHz, 1 MHz and 2 MHz is predominantly driven by non-thermal, optical forces.

\begin{figure}
\includegraphics[width=80mm]{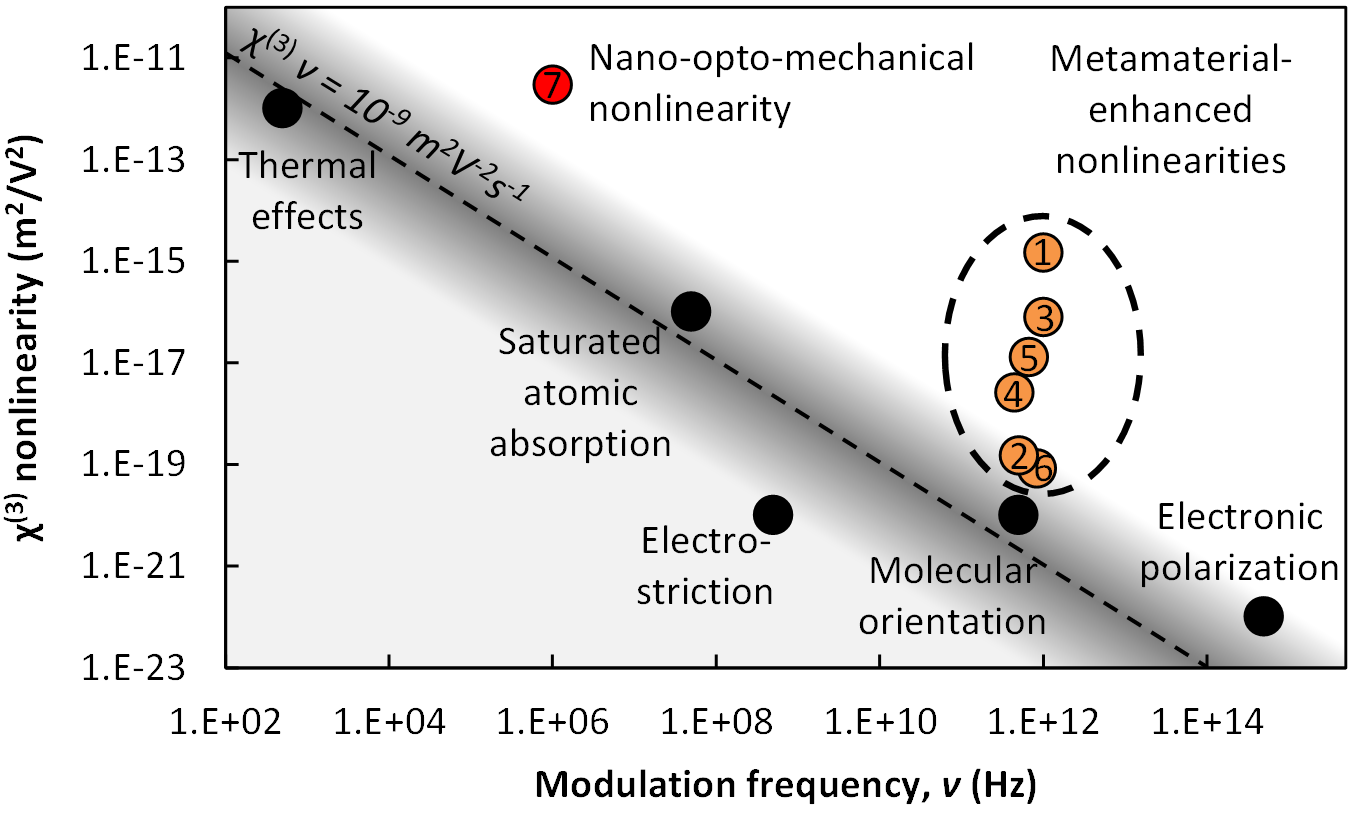}
\caption{\label{fig-nonlinearity}
\textbf{Optical nonlinearities in metamaterials and conventional media.} Cubic optical nonlinearities due to thermal effects, saturated absorption, molecular orientation and electronic anharmonicity follow a general trend of decreasing achievable modulation frequency $\nu$ with increasing magnitude of the nonlinearity: $\chi^{(3)} \nu \sim 10^{-9} \text{m}^2 \text{V}^{-2} \text{s}^{-1}$ \cite{Boyd_NLO_3rdEd_p211}. Metamaterial patterning of plasmonic metal (points 1 and 2, \cite{RenNonlinearGold, wurtz2011ultrafastNL}), hybridization of carbon nanotubes (point 3, \cite{OE_Nonlinear_CNT_MM_Nikolaenko}), graphene (point 4, \cite{APL_2012_nlGraphene}), and semiconductors (points 5 and 6, \cite{cho2009ultrafast, Dani}) with metamaterials resonantly enhances the magnitude of the nonlinear response without affecting its response time. Nano-opto-mechanical metamaterial demonstrated in the present work also departs from the common trend exhibiting $\chi^{(3)} \nu$ that is three orders of magnitude bigger than in conventional nonlinear media (point 7).
}
\end{figure}

Absorption in a nonlinear medium is conventionally described by the expression $-\tfrac{dI}{dz}=\alpha I + \beta I^2 + ...$, where $I$ is light intensity, $z$ is the propagation distance and $\alpha$ and $\beta$ are the linear and nonlinear absorption coefficients. As the observed nonlinear transmission change is proportional to the pump power, we can quantify the nonlinearity of the reconfigurable photonic metamaterial by estimating its first nonlinear absorption coefficient $\beta$.
Assuming that the nonlinear transmission change $\Delta T$ results from nonlinear absorption, 
$\beta\sim\Delta T/(It)$, where $t$ is the metamaterial's thickness. 
At 1~MHz modulation frequency, $\beta\sim5\cdot 10^{-3}$~m/W corresponding to a nonlinear susceptibility on the order of $\text{Im}(\chi^{(3)})/n^2\sim 10^{-12} \text{m}^2 \text{V}^{-2}$ and a nonlinear imaginary refractive index change $\Delta \kappa\sim 5\cdot 10^{-3}$ at 920~W/cm$^2$ pump intensity.
We argue here that the observed nano-opto-mechanical nonlinearity is special as it exhibits a departure from the prevailing relation between speed and magnitude of nonlinear responses (Fig.~\ref{fig-nonlinearity}). This happens in a way that allows larger nonlinearity or higher modulation frequency to be achieved than could be expected from traditional media. Indeed, it appears that for many mechanisms of optical nonlinearity the achievable modulation frequency $\nu$ is related to the magnitude of the nonlinerity $\chi^{(3)}$ in such a way that $\chi^{(3)} \nu \sim 10^{-9} \text{m}^2 \text{V}^{-2} \text{s}^{-1}$ \cite{Boyd_NLO_3rdEd_p211}. At the resonance of in-plane motion, MHz rate modulation can be achieved at about 1~kW/cm$^2$ of light intensity corresponding to about $\chi^{(3)} \nu \sim 10^{-6} \text{m}^2 \text{V}^{-2} \text{s}^{-1}$.

In summary, we provide the first experimental demonstration of a giant nano-opto-mechanical nonlinearity in plasmonic metamaterial that allows modulation of light with light at MHz frequencies and miliwatt power levels. Bimorph deformation resulting from optical heating at low modulation frequencies and electromagnetic forces between elements of plasmonic resonators drive the nonlinear response. The magnitude and spectral position of the underlying optical resonance can be controlled across the optical spectral range by design of the plasmonic resonators and the supporting nano-membrane strips.

\section{Methods}

\subsection{Plasmonmechanical metamaterial fabrication.}

Starting with a commercially available 50~nm thick low stress silicon nitride membrane, a 50~nm thick gold layer for the plasmonic metamaterial was thermally evaporated. The gold-coated membrane was structured with a focussed ion beam system (FEI Helios 600 NanoLab). The pattern of plasmonic resonators was milled and then the supporting silicon nitride membrane was cut into suspended strips with tapered ends. Detailed dimensions of the nanostructure are given by Fig.~\ref{fig-sample}b.

\subsection{Experimental characterization.}

The cubic nano-opto-mechanical nonlinearity was measured by detecting the pump-induced modulation of the metamaterial's transmission (Fig.~\ref{fig-modulation}a). The probe beam at wavelength 1310~nm was generated by a fiber-coupled telecom laser diode (Thorlabs FPL1053S). The pump beam from a fiber-coupled laser diode at wavelength 1550~nm (Thorlabs FPL1009S) was modulated with a fiberized electro-optical modulator (EOSpace AX-0K5-10-PFA-UL) and then combined with the probe beam in a fiber coupler (JDSU L2SWM1310/1550BB). The transmitted beam was filtered to remove the pump and the probe signal was detected by an InGaAs photodetector (New Focus 1811) and a lock-in amplifier (Stanford Research SR844).

In all optical experiments and simulations the electric field was $x$-polarized (parallel to the strips) and incident on the silicon nitride side of the sample.

\section{Acknowledgements}

The authors are grateful to Artemios Karvounis for fruitful discussions and to Pablo Cencillo-Abad for assistance in preparing the manuscript.
This work is supported by the Leverhulme Trust, the
Royal Society, the U.S. Office of Naval Research (grant N000141110474), the MOE Singapore (grant MOE2011-T3-1-005) and the UK's Engineering and Physical Sciences
Research Council (grant EP/G060363/1). The data from this paper can be obtained from the University of Southampton ePrints research repository: http://dx.doi.org/10.5258/SOTON/377861

\bibliographystyle{nature}

\end{document}